\documentclass[fleqn,usenatbib]{mnras}

\usepackage{newtxtext,newtxmath}

\usepackage[T1]{fontenc}
\usepackage{ae,aecompl}
\usepackage{etoolbox}
\makeatletter
\patchcmd\@combinedblfloats{\box\@outputbox}{\unvbox\@outputbox}{}{\errmessage{\noexpand patch failed}}
\makeatother

\usepackage{graphicx}	
\usepackage{amsmath}	
\usepackage{amssymb}	
\usepackage{epsfig}
\usepackage{xcolor}

\newcommand{\grbs}{gamma-ray bursts}
\newcommand{\ngrbs}{20}
\newcommand{\GRB}{GRB}
\newcommand{\GRBs}{GRBs}
\newcommand{\FRB}{FRB}
\newcommand{\FRBs}{FRBs}
\newcommand{\craft}{CRAFT}
\newcommand{\askap}{ASKAP}
\newcommand{\obsstartdate}{2018 March 11}
\newcommand{\obsenddate}{2018 July 3}
\newcommand{\ntriggers}{29}

\date{}

\pubyear{2019}

\title[A search for \FRB\ emission from \textit{Fermi} \GRBs]{A search for fast radio burst-like emission from \textit{Fermi} \grbs}
\author[Bouwhuis et al.]{Mieke~Bouwhuis$^{1,2}$\thanks{\href{mailto:mieke.bouwhuis@csiro.au}{mieke.bouwhuis@csiro.au}},
Keith~W.~Bannister$^{1}$,
Jean-Pierre~Macquart$^{3}$,
R.~M.~Shannon$^{4}$, \newauthor
David~L.~Kaplan$^{5}$,
John~D.~Bunton$^{1}$,
B\"arbel~S.~Koribalski$^{1}$,
M.~T.~Whiting$^{1}$
\\
$^{1}$CSIRO Astronomy and Space Science (CASS), P.O. Box 76, Epping, NSW 1710, Australia\\
$^{2}$NIKHEF, National Institute for Subatomic Physics, Amsterdam, The Netherlands \\
$^{3}$International Centre for Radio Astronomy Research, Curtin University, Bentley, WA, Australia \\
$^{4}$Centre for Astrophysics and Supercomputing, Swinburne University of Technology, Hawthorn  VIC 3122, Australia \\
$^{5}$Center for Gravitation, Cosmology, and Astrophysics, Department of Physics, University of Wisconsin-Milwaukee, P.O. Box 413,\\ Milwaukee, WI 53201, USA
}

\begin{document}
\label{firstpage}
\pagerange{\pageref{firstpage}--\pageref{lastpage}}
\maketitle

\begin{abstract}
We report the results of the rapid follow-up observations of \grbs\ (\GRBs) detected by the \textit{Fermi} satellite to search for associated fast radio bursts.  The observations were conducted with the Australian Square Kilometre Array Pathfinder at frequencies from 1.2--1.4 GHz.  
A set of 20 bursts, of which four were short \GRBs, were followed up with a typical latency of about one minute, for a duration of up to 11 hours after the burst. 
The data was searched using 4096 dispersion measure trials up to a maximum dispersion measure of 3763\,pc\,cm$^{-3}$, and for pulse widths $w$ over a range
of duration from 1.256 to 40.48\,ms.
No associated pulsed radio emission was observed above $26 \, {\rm Jy\,ms}\, (w/1\,{\rm ms})^{-1/2}$ for any of the \ngrbs\ \GRBs.
\end{abstract}


\section{Introduction}
The origin and source of fast radio bursts~\citep[\FRBs;][]{Lorimeretal2007} are still unknown. 
There are a number of theoretical models that attempt to connect short \grbs\ (\GRBs), or more specifically binary neutron star mergers, with fast radio burst-like phenomena~\citep[e.g.,][]{UsovKatz2000,PshirkovPostnov2010,Totani2013,FalckeRezzolla2014,RaviLasky2014,Zhang2014,Wangetal2016,Wangetal2018,Lyutikov2018}. The overall fast radio burst (\FRB) detection rate precludes short \GRBs\ as the only channel for \FRB\ emission, as discussed in~\citet{Ravi2019}.
Nonetheless, since short \GRBs\ are generally believed to originate from neutron star--neutron star mergers~\citep{Gehrels2005,Berger2014,Fong2015} the follow-up of gamma-ray burst (\GRB) detections coincident with impulsive radio emission provides a means to test whether such mergers can generate some fraction of the total \FRB\ population.

While the majority of theoretical work has concentrated on short \GRBs, the~\citet{UsovKatz2000} model could also be applied to long \GRBs, 
but in that case it is difficult to understand how the radio emission could propagate through the extended stellar envelope~\citep{Macquart2007}.  

Many of the models predict \FRB-like radio emission just prior to, or coincident with the merger, when the gamma rays are released.
Unambiguous identification of this `prompt' \FRB\ emission requires simultaneous observations in gamma rays and radio, which is difficult to achieve in practice. 

The fact that radio emission is delayed due to cold plasma dispersion does not help when observing at $\sim$1~GHz (where most \FRBs\ have been found to date).
For a typical GRB redshift of $z\sim1$~\citep{Gruberetal2014} the expected dispersion measure from intergalacic cold plasma is 
$\sim\,10^3$~pc\,cm$^{-3}$~\citep{Ioka2003,Inoue2004}, leading to a dispersion delay of only $\sim$2~s, which is insufficient time to slew mechanically steered telescopes.

Nevertheless, the \citet{RaviLasky2014} model \emph{does} predict `delayed' \FRB-like emission, $10$--$10^4$~s after the merger, well within the reaction time of current facilities.
In this picture, the merger of two neutron stars creates a `supramassive' millisecond neutron star, which subsequently collapses into a black hole.
This collapse may be observable if it releases \FRB-like emission. 

Here we present the results of a search for delayed, \FRB-like radio emission associated with known short \GRB\ events, specifically aiming to test the \citet{RaviLasky2014} proposal.

A limiting factor in observing the prompt emission of short \GRBs\ in particular is their relatively poor localization.
Hence an instrument with a large field of view is required.
In addition, a short slewing time is required to probe the case where the collapse follows the merger very quickly.
The follow-up observations were performed with the Australian Square Kilometre Array Pathfinder \citep[\askap;][]{McConnell2016,Hotan2014} as part of the Commensal Real-time \askap\ Fast Transients (\craft) survey~\citep{Macquart2010}, because of its large field of view and rapid follow-up capabilities.

Previous searches for \FRB\ emission at $\sim$1~GHz~\citep{Bannister2012, Palaniswamy2014}, where most \FRBs\ have been detected so far, required good localisation and hence used \GRBs\ that were detected by the \textit{Swift} satellite~\citep{Gehrels2004}, which are generally long \GRBs.
\citet{Kaplan2015} on the other hand followed up a short \textit{Swift} \GRB\ at 132~MHz, but no \FRBs\ have been observed at this frequency~\citep{Sokolowski2018}.  
The key feature of the study presented here comes from the fact that we observe short \GRBs\ detected by the \textit{Fermi} satellite~\citep{Meegan2009} at a frequency where \FRBs\ are known to occur and we can cover the \textit{Fermi} localisation uncertainty because of the large field of view of \askap.

The paper is partitioned as follows.
We detail the data acquisition and processing in Section \ref{sec:obs}, present our results in Section \ref{sec:res}, discuss the implications of these findings in Section \ref{sec:discuss} and conclude in Section \ref{sec:conclusion}.

\section{Data acquisition and processing}
\label{sec:obs}
A significant challenge in the search for prompt radio emission from \GRBs\ is that their initial detection positions are not localised to regions smaller than tens of square degrees.
This mandates an approach which is able to instantaneously cover as much of the initial localisation region as possible.
To this end we configured the antennas of \askap\ into a fly's-eye configuration, similar to the arrangement described in~\citet{Bannister2017}.
We observed at a central frequency of 1296~MHz. 
Each antenna has 36 beams covering 30~square\,degrees field of view in a hexagonally close-packed arrangement, with a pitch equal to the beam full-width-half-maximum at the center of the band of 0.9~degrees.
The observations were taken during commissioning mode, in which typically only 6-8 antennas were available, yielding an instantaneous field of view between 180 and 240~square\,degrees. 

For details of the recording and search pipelines, we refer the reader to \citet{Bannister2017}, but we briefly describe the main points below.
Data were recorded using the the real-time \craft\ data pipeline as described in~\citet{Clarke2014}, which records 336$\times$1\,MHz channel Stokes-I autocorrelations with a time resolution of 1.265~ms.
The offline \FRB\ search was done with the GPU-based \FRB\ detection pipeline called Fast Real-time Engine for Dedispersing Amplitudes (FREDDA).
The main purpose of FREDDA is to take into account the effect of dispersion.
Any impulsive radio signal propagating across cosmological distances is strongly affected by dispersion caused by plasma in our Galaxy, the intergalactic medium and the burst host galaxy.  The retardation in the pulse arrival time varies quadratically with wavelength and scales linearly with the total electron column density, the dispersion measure (DM). The DM of any putative radio burst is unknown, so the signal must be dedispersed and searched over a grid of plausible DM values.
The FREDDA algorithm searched for pulses using 4096 DM trials up to a maximum DM of 3763\,pc\,cm$^{-3}$, with a resolution of 0.92\,pc\,cm$^{-3}$, and for pulse widths $w = T, 2T, 3T ... 32T$ where the integration time was $T=1.265$\,ms.
The \FRB\ candidates that were found and saved to disk by the \FRB\ search algorithm FREDDA were visually inspected by looking at the shape of the pulse as a function of time and frequency, and the signal-to-noise ratio.

\subsection{\GRB\ follow-up observations}
In the search for \FRBs\ associated with short \GRBs\ we focused on the \GRBs\ detected 
by the GBM instrument on the \textit{Fermi} satellite~\citep{Goldstein2012}.
The GBM instrument has the highest detection rate of short \GRBs\ of about 40--45 per year.
A \textit{Fermi} \GRB\ detection is followed by the distribution of different types of alerts to allow for follow-up observations.
The only type of follow-up alert messages responded to by \askap\ were those distributed with a relatively short delay after detection, and with a sufficiently accurate position to permit radio follow-up.
These message types include the so-called \textit{GBM Ground position notices} and \textit{GBM Final position notices}.
The former contain the position of the source with an uncertainty of about 1--10 degrees radius and are distributed within a few tens of seconds.
The latter are distributed with a much larger delay, of up to a few hours, but contain an updated source position with a smaller uncertainty of about 5~degrees radius.

\begin{figure*}
\begin{center}
\includegraphics{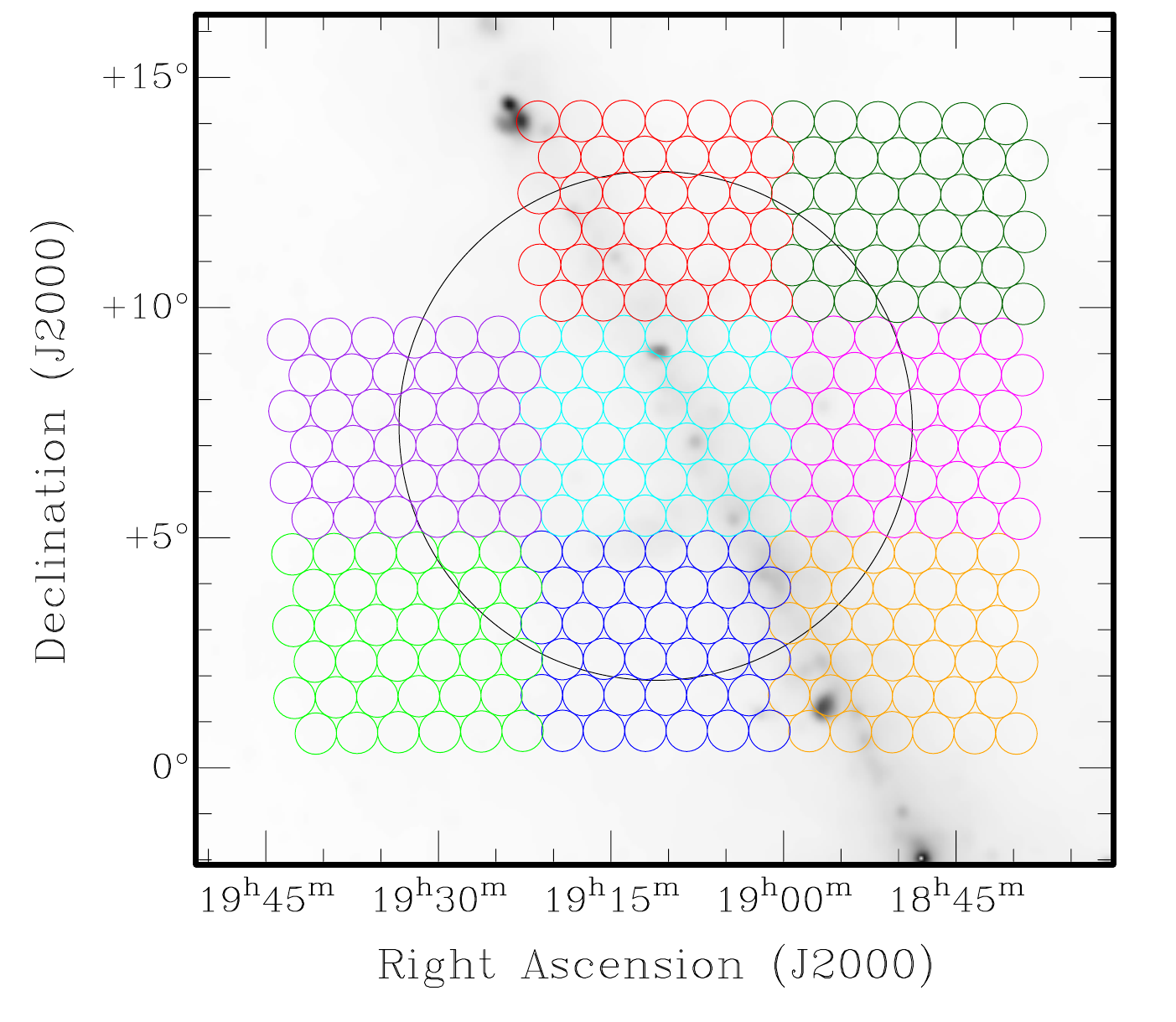}
\caption{The tiling of 8 antenna footprints on the sky for the follow-up observation of the \GRB\,180610B detection by \textit{Fermi}. 
Each set of 36~beams in the same color represents one \askap\ antenna. The footprint in the middle is one antenna that is pointed in the direction of the source position provided in the \GRB\ alert message. 
The black circle indicates the uncertainty of 5.53 degrees radius on the source position provided in the \GRB\ alert message. 
This uncertainty is the circular area equivalent of the statistical uncertainty \citep[68\% confidence level;][]{Narayana2016}.
The background shows the radio continuum map published in~\citet{Calabrettaetal2014}.}
\label{fig:tilesky}
\end{center}
\end{figure*}

The 4PiSky~\citep{Staley2016} package was used to handle the \GRB\ alerts.
As we were interested in low latency \GRB\ follow-up, \askap\ observations were only triggered in response to a \GRB\ detection if the reported source position was above the horizon at the time.
The footprints of the antennas on the sky were then tiled in a fly's-eye configuration, with each telescope pointing in a different direction,
to cover the uncertainty on the source location, as shown in Figure~\ref{fig:tilesky}.
For the follow-up observation all antennas that were available for the \craft\ survey were used.

In the tiling of the localisation region (as shown in Figure~\ref{fig:tilesky}), the centre of one antenna is pointed at the position that is given in the \GRB\ alert message.
The remaining antennas are pointed as close as possible to the direction of this central antenna without the footprints overlapping each other.
For each new alert message corresponding to the same burst detection the pointing of all antennas
is adjusted to the updated source location given in the alert message.
The follow-up observation interrupts the ongoing \craft\ observation and it lasts in principle until the source sets.

\subsection{\GRB\ follow-up details}
The \GRBs\ that were followed up were detected between \obsstartdate\ and \obsenddate.
A total of \ntriggers\ GCN alerts~\citep{Barthelmy1994} were handled in this period of which 
\ngrbs\ are classified as \GRBs, listed in Table~\ref{table:grbs}.
All \GRBs\ were detected by the \textit{Fermi} satellite except for one.
\GRB\,180331A was detected by the \textit{Swift} satellite on which \askap\ triggered during a period
where \textit{Fermi} was not operational.

\begin{table*}
\centering
\begin{tabular}{lrrrrrrr}
\hline
\GRB\ name		& Trigger ID & \GRB\ time & \GRB\ duration & Fluence & Error circle &  On-source	&       Total observa- \\
                    &			 & [UT]       &   [s] & [erg\,cm$^{-2}$] & coverage & delay [s] &  tion time [h]  \\
\hline

\GRB\,180313A*	& bn180313978 & 23:28:17.53 	& 0.080  & 1.8732\,x\,10$^{-7}$ &  0.86 & 147           &       5.3     \\

\GRB\,180331A	& 820347      & 04:14:55.70 	&48.0 & 6.1\,x\,10$^{-7}$ &  1.00 &   104           &       7.1     \\ 
\GRB\,180404B	& bn180404091 & 02:11:38.64 	&  80.897 & 2.8009\,x\,10$^{-5}$ & 1.00 & 148           &       9.7     \\

\GRB\,180412A	& bn180412425 & 10:12:06.01	&  20.992 & 2.9571\,x\,10$^{-6}$ & 1.00 &67        &       3.3 \\ 

\GRB\,180416A	& bn180416340 & 08:09:26.47	&   103.426 & 3.8795\,x\,10$^{-5}$ & 1.00 &111           &       6.7     \\

\GRB\,180420B	& bn180420107 & 02:33:54.63	&  62.977 & 3.0334\,x\,10$^{-6}$ & 0.99 & 134           &       11.2    \\

\GRB\,180423B	& bn180423266 & 06:23:40.76	&  3.328 & 5.7854\,x\,10$^{-7}$ & 0.35 & 133           &       7.8     \\

\GRB\,180511B*	& bn180511437 & 10:29:52.61	&   1.984 & 7.6532\,x\,10$^{-7}$ & 0.64 &128           &       2.3     \\

\GRB\,180511C	& bn180511606 & 14:32:18.99	&   9.216 & 1.3554\,x\,10$^{-6}$  & 0.88 &84            &       6.0     \\

\GRB\,180513A	& bn180513815 & 19:34:02.14	&   20.480 & 1.6211\,x\,10$^{-6}$ & 0.93 &118           &       1.0     \\

\GRB\,180516A	& bn180516229 & 05:29:47.14	&  13.824 & 1.7834\,x\,10$^{-6}$ & 0.95 &143           &       0.2     \\

\GRB\,180521A	& bn180521935 & 22:26:57.24	&  31.232  & 6.6444\,x\,10$^{-7}$ & 0.84 &101           &       4.0     \\

\GRB\,180525A*	& bn180525151 & 03:37:59.08	&   0.544 & 1.4069\,x\,10$^{-7}$ & 0.26 &135           &       7.3     \\

\GRB\,180528B	& bn180528465 & 11:09:49.03	&   8.192 & 5.4004\,x\,10$^{-7}$ & 0.28 &135           &       6.2     \\

\GRB\,180602B*	& bn180602938 & 22:31:05.34	&   0.008 & 2.0878\,x\,10$^{-7}$ & 1.00 &145           &       7.4     \\

\GRB\,180606A	& bn180606730 & 17:31:38.62	&   6.080 & 1.2997\,x\,10$^{-6}$  & 0.12 &124           &       5.3     \\

\GRB\,180610A	& bn180610377 & 09:03:13.18	&   163.331 & 8.5794\,x\,10$^{-6}$ & 0.94 &65            &       4.6     \\

\GRB\,180610B	& bn180610568 & 13:38:27.51	&  8.448 & 1.1805\,x\,10$^{-6}$ & 0.96 &123           &       5.3     \\

\GRB\,180610C    & bn180610791 & 18:58:48.67	&   27.648 & 4.0315\,x\,10$^{-6}$ & 1.00 &89            &       3.2     \\

\GRB\,180618B	& bn180618724 & 17:22:06.25	&   130.050 & 1.8649\,x\,10$^{-5}$  & 0.94 &100           &       1.0     \\

\hline
\end{tabular}
\caption{The name, trigger ID, time, duration, and fluence of the \GRBs\ that were followed up by an \askap\ observation. 
Also indicated are the error circle coverage, which is the fraction of the 68\% confidence region that was covered by the \askap\ dishes, the on-source delay, which is the time it took to be on source (in seconds), and the total follow-up observation time (in hours).
\GRBs\ marked with an * are classified as short duration \GRBs. 
\GRB\,80331A was detected by the \textit{Swift} satellite in the period that \textit{Fermi} was not operational. All other \GRBs\ were detected by the \textit{Fermi} satellite. }
\label{table:grbs}
\end{table*}

No redshift information is available in any of the corresponding GCN circulars\footnote{\url{https://gcn.gsfc.nasa.gov/gcn3\_circulars.html}} for the \GRBs\ listed.
None of the \GRBs\ listed was localized by the IPN\footnote{\url{https://heasarc.gsfc.nasa.gov/W3Browse/all/ipngrb.html}}, and no associated sources were found in the TNS\footnote{\url{https://wis-tns.weizmann.ac.il/}} archive.

Apart from the \GRB\ specifications, Table~\ref{table:grbs} also lists the relevant \askap\ observation parameters: the error circle coverage, the on-source delay, and the total observation time.
The error circle coverage indicates the fraction of the error on the \GRB\ position that is covered by the \askap\ dishes for the corresponding observation.
The error circle coverage was calculated using the uncertainty in the \GRB\ location given in the last alert message for a certain \GRB\ detection that the system triggered on.
For most \textit{Fermi} \GRBs\ this was the position uncertainty given in the \textit{GBM Final position notice}, except for \GRB\,180313A, \GRB\,180513A, and \GRB\,180521A, for which the position uncertainty in the \textit{GBM Ground position notice} was used. 
Our radio follow-up observations covered, on average, 80\% of the 68\% confidence containment region.
The on-source delay, defined as the delay between the \GRB\ time and the start of the observation by \askap, was on average about 120~s.
The total observation time, which is the duration of the follow-up observation by \askap, was on average about 5.3 hours.

Four of the \GRBs\ in Table~\ref{table:grbs} are classified as short \GRBs, which is consistent with the fraction of about 21\% of short \GRBs\ in the \textit{Fermi} GBM sample \GRBs~\citep{Narayana2016}.  

\section{Results} \label{sec:res}

We have demonstrated the prompt follow-up of \ngrbs\ \GRBs\ with the \askap\ radio telescope. 
Four of these \GRBs\ are classified as short \GRBs.
The observations started 65--148~s after the burst, and lasted for a few hours, up to 11.2 hours after the detection of the burst.

A total of 433 single \FRB\ candidates were output by the search pipeline above a signal-to-noise ratio of 10 during the course of the follow-up observations, with the number of events found per observation ranging from 0 to 88.
All candidates were visually inspected, and were readily attributable to radio frequency interference or known instrumental artifacts.
In addition, for the follow-up of \GRB\,180416A the bright pulsar PSR B0833$-$45 was in the field of view and generated many candidates. This pulsar detection demonstrates that the acquisition and search pipelines were operating correctly and at the correct sensitivity.
No viable \FRB\ candidates were found during the observations reported here.

Based on the sensitivity of the \craft\ survey to \FRBs~\citep{James2019}, the upper limit of burst-like
radio emission from \GRBs\ after about 120~s, and within a few hours, after the detection of the burst is
$26 \, {\rm Jy\, ms}\, (w/1\,{\rm ms})^{-1/2}$, where the integration time $w$ ranges from
\mbox{1.265--40.48}~ms.
The median efficiency of \craft\ fly's-eye observations has been estimated at 90\%~\citep{James2019}, i.e. there is a 10\% chance of an \FRB\ being missed in the search pipeline, e.g. due to radio frequency interference or data losses.
This number reflects data taken during commissioning, and thus represents an upper bound to the probability of missing an \FRB.
The sensitivity varies slightly between telescope and the beam pattern, and with increasing DM.
The quoted number is the sensitivity in the middle of the central beams only.

\section{Discussion}
\label{sec:discuss}
Possible explanations for this null result include a mismatch between any prompt emission and the follow-up timescales, the finite sensitivity of the follow-up, and the physical absence of FRB-like radio emission.  We discuss each of these below.

Although the arrival of any radio emission would be retarded relative to the gamma-ray emission due to plasma dispersion along the burst sight line (likely chiefly due to the intergalactic medium), we note that this effect is too small to enable us to observe radio emission that would have coincided with the \GRB\ itself.  
The total dispersive delay for any event with redshift $z<3$ is typically less than 6\,s, at an observing frequency of 1300\,MHz, far less than the average \askap\ response time of about 120~s.
Moreover, at a typical \GRB\ redshift of $z\sim1$~\citep{Gruberetal2014}, the DM is expected to be $\sim 900\,$pc\,cm$^{-3}$~\citep{Ioka2003,Inoue2004}, which corresponds to a dispersive time delay of only $\sim\,2$~s.
Due to the instrumental capabilities we are thus unable to test models that predict `precursor' or `prompt' \FRB-like emission from \GRBs.

Our observations do, however, direct address the proposal of~\citet{RaviLasky2014}, wherein a pair of merged neutron stars may persist as an unstable supramassive neutron star for up to a few hours after merger, and thus for which post-merger emission may be expected.
The optimal detection window for radio emission in this model is between 10~s and $4.4\times10^4$~s after the initial burst.
We observed 4 short \GRBs\ within about $2\times10^2$--$2\times10^4$~s after the \GRB\ time.
It is possible that the collapse occurred outside our observing window. 
To investigate the likelihood, we integrated under the curves shown in Figure 3 of~\citet{RaviLasky2014}. 
We find that our average observation, which covered a collapse time of roughly $2\times10^2<t<2\times10^4$~s, covers 39\%, 56\% and 40\% of the predicted collapse time distributions for the AB-N, GM1 and APR models respectively. 
Therefore, there is a reasonable chance of all 4 \GRBs\ collapsing outside the observing window. 

A second explanation is insufficient sensitivity.
Our fluence sensitivity is $26 \, {\rm Jy\, ms}\, (w/1\,{\rm ms})^{-1/2}$. 
Assuming a fiducial redshift for short \GRBs\ of $z\sim0.85$~\citep{Gruberetal2014}, this corresponds to an energy spectral density of $7 \times 10^{25} (w/1\,{\rm ms})^{-1/2}$\,J\,Hz$^{-1}$ (assuming isotropic emission and a flat radio spectrum), a value which is intermediate amongst the population of FRBs currently observed \citep[see figure 2 of][]{Shannon18}.

Thirdly, it is possible that the progenitor of these short \GRBs\ did not produce any \FRBs\ at all.
\citet{RaviLasky2014} state in Section 4 of their paper that the probability of binary neutron star mergers to result in supramassive stars ranges between 5\%--95\%, depending on the model. 
This would mean that for the four short \GRBs\ in our data sample, this could have lead to the detection of zero to four \FRB\ events.
The dynamic range of the models is such that with the null observation the binary star merger resulting in a supramassive star can not be concluded. 
On the other hand,~\citet{RaviLasky2014} also state that, based on the short \GRB\ X-ray lightcurve sample of~\citet{Rowlinson2013}, it is likely that between 15\% and 25\% of short \GRBs\ will result in supramassive stars.
This means that zero or one of the short \GRBs\ in our data sample could have lead to an \FRB.
Again, the current statistics are not conclusive.

On average we covered about 70\% of the uncertainty region for these \GRBs.
We estimate the probability of all 4 \GRBs\ falling outside our coverage as $<$\,0.05\%.
We observed at a frequency of 1296~MHz (where \FRBs\ are known to occur) and with sufficient time and frequency resolution to detect short bursts.
Overall we find that our observations were sufficient to detect any \FRBs\ emitted by the 4 short \GRBs\ observed, had they been sufficiently bright and had the collapse occurred while the observations were taking place.

\section{Conclusion} \label{sec:conclusion}
We have searched for pulsed radio emission from \GRBs\ at a central frequency of 1296~MHz.
Similar follow-up observations to detect \FRBs\ in coincidence with the prompt emission of \GRBs\ in general have been performed before~\citep{Bannister2012, Palaniswamy2014, Kaplan2015}, however this search focused on short \GRBs, which are currently primarily detected by the \textit{Fermi} satellite, at a frequency where \FRBs\ are known to occur. 
\askap\ with its large field of view and fast slewing time is currently the most capable radio telescope to perform these observations.
We observed \ngrbs\ \GRBs\ by quickly responding to the GCN notices and pointing the \askap\ dishes in the direction of the source within 65--148 seconds.
We have found no \FRB\ events collected during these observations.
The $\sim120$~s delay between the detection of the \GRB\ and the start of the radio observations limits the conclusions we are able to draw on most of the proposed models, except notably that of~\citet{RaviLasky2014}.

\section*{Data availability}
Data are archived at the Pawsey Supercomputing Centre. 
The authors will attempt to provide reasonable requests for data.

\section*{Acknowledgements}
    The Australian SKA Pathfinder is part of the Australia Telescope National Facility which is managed by CSIRO. Operation of \askap\ is funded by the Australian Government with support from the National Collaborative Research Infrastructure Strategy. \askap\ uses the resources of the Pawsey Supercomputing Centre. Establishment of \askap, the Murchison Radio-astronomy Observatory and the Pawsey Supercomputing Centre are initiatives of the Australian Government, with support from the Government of Western Australia and the Science and Industry Endowment Fund. We acknowledge the Wajarri Yamatji people as the traditional owners of the Observatory site. DK was supported by NSF grant AST-1816492.

\bibliographystyle{mnras}
\bibliography{references-GRBFRB}

\begin{thebibliography}{}
\makeatletter
\relax
\def\mn@urlcharsother{\let\do\@makeother \do\$\do\&\do\#\do\^\do\_\do\%\do\~}
\def\mn@doi{\begingroup\mn@urlcharsother \@ifnextchar [ {\mn@doi@}
  {\mn@doi@[]}}
\def\mn@doi@[#1]#2{\def\@tempa{#1}\ifx\@tempa\@empty \href
  {http://dx.doi.org/#2} {doi:#2}\else \href {http://dx.doi.org/#2} {#1}\fi
  \endgroup}
\def\mn@eprint#1#2{\mn@eprint@#1:#2::\@nil}
\def\mn@eprint@arXiv#1{\href {http://arxiv.org/abs/#1} {{\tt arXiv:#1}}}
\def\mn@eprint@dblp#1{\href {http://dblp.uni-trier.de/rec/bibtex/#1.xml}
  {dblp:#1}}
\def\mn@eprint@#1:#2:#3:#4\@nil{\def\@tempa {#1}\def\@tempb {#2}\def\@tempc
  {#3}\ifx \@tempc \@empty \let \@tempc \@tempb \let \@tempb \@tempa \fi \ifx
  \@tempb \@empty \def\@tempb {arXiv}\fi \@ifundefined
  {mn@eprint@\@tempb}{\@tempb:\@tempc}{\expandafter \expandafter \csname
  mn@eprint@\@tempb\endcsname \expandafter{\@tempc}}}

\bibitem[\protect\citeauthoryear{{Bannister}, {Murphy}, {Gaensler}  \&
  {Reynolds}}{{Bannister} et~al.}{2012}]{Bannister2012}
{Bannister} K.~W.,  {Murphy} T.,  {Gaensler} B.~M.,   {Reynolds} J.~E.,  2012,
  \mn@doi [\apj] {10.1088/0004-637X/757/1/38}, \href
  {https://ui.adsabs.harvard.edu/abs/2012ApJ...757...38B} {757, 38}

\bibitem[\protect\citeauthoryear{{Bannister} et~al.,}{{Bannister}
  et~al.}{2017}]{Bannister2017}
{Bannister} K.~W.,  et~al., 2017, \mn@doi [\apjl] {10.3847/2041-8213/aa71ff},
  \href {http://cdsads.u-strasbg.fr/abs/2017ApJ...841L..12B} {841, L12}

\bibitem[\protect\citeauthoryear{{Barthelmy} et~al.,}{{Barthelmy}
  et~al.}{1994}]{Barthelmy1994}
{Barthelmy} S.~D.,  et~al., 1994, in {Fishman} G.~J.,  ed.,  American Institute
  of Physics Conference Series Vol. 307, Gamma-Ray Bursts. p.~643,
  \mn@doi{10.1063/1.45819}

\bibitem[\protect\citeauthoryear{{Berger}}{{Berger}}{2014}]{Berger2014}
{Berger} E.,  2014, \mn@doi [\araa] {10.1146/annurev-astro-081913-035926},
  \href {http://adsabs.harvard.edu/abs/2014ARA%26A..52...43B} {52, 43}

\bibitem[\protect\citeauthoryear{{Calabretta}, {Staveley-Smith}  \&
  {Barnes}}{{Calabretta} et~al.}{2014}]{Calabrettaetal2014}
{Calabretta} M.~R.,  {Staveley-Smith} L.,   {Barnes} D.~G.,  2014, \mn@doi
  [Publications of the Astronomical Society of Australia]
  {10.1017/pasa.2013.36}, \href
  {https://ui.adsabs.harvard.edu/#abs/2014PASA...31....7C} {31, e007}

\bibitem[\protect\citeauthoryear{{Clarke}, {D'Addario}, {Navarro}  \&
  {Trinh}}{{Clarke} et~al.}{2014}]{Clarke2014}
{Clarke} N.,  {D'Addario} L.,  {Navarro} R.,   {Trinh} J.,  2014, \mn@doi
  [Journal of Astronomical Instrumentation] {10.1142/S2251171714500044}, \href
  {http://cdsads.u-strasbg.fr/abs/2014JAI.....350004C} {3, 1450004}

\bibitem[\protect\citeauthoryear{{Falcke} \& {Rezzolla}}{{Falcke} \&
  {Rezzolla}}{2014}]{FalckeRezzolla2014}
{Falcke} H.,  {Rezzolla} L.,  2014, \mn@doi [\aap]
  {10.1051/0004-6361/201321996}, \href
  {http://adsabs.harvard.edu/abs/2014A%26A...562A.137F} {562, A137}

\bibitem[\protect\citeauthoryear{{Fong}, {Berger}, {Margutti}  \&
  {Zauderer}}{{Fong} et~al.}{2015}]{Fong2015}
{Fong} W.,  {Berger} E.,  {Margutti} R.,   {Zauderer} B.~A.,  2015, \mn@doi
  [\apj] {10.1088/0004-637X/815/2/102}, \href
  {http://adsabs.harvard.edu/abs/2015ApJ...815..102F} {815, 102}

\bibitem[\protect\citeauthoryear{{Gehrels} et~al.,}{{Gehrels}
  et~al.}{2004}]{Gehrels2004}
{Gehrels} N.,  et~al., 2004, \mn@doi [\apj] {10.1086/422091}, \href
  {https://ui.adsabs.harvard.edu/abs/2004ApJ...611.1005G} {611, 1005}

\bibitem[\protect\citeauthoryear{{Gehrels} et~al.,}{{Gehrels}
  et~al.}{2005}]{Gehrels2005}
{Gehrels} N.,  et~al., 2005, \mn@doi [\nat] {10.1038/nature04142}, \href
  {https://ui.adsabs.harvard.edu/abs/2005Natur.437..851G} {437, 851}

\bibitem[\protect\citeauthoryear{{Goldstein} et~al.,}{{Goldstein}
  et~al.}{2012}]{Goldstein2012}
{Goldstein} A.,  et~al., 2012, \mn@doi [\apjs] {10.1088/0067-0049/199/1/19},
  \href {http://adsabs.harvard.edu/abs/2012ApJS..199...19G} {199, 19}

\bibitem[\protect\citeauthoryear{{Gruber} et~al.,}{{Gruber}
  et~al.}{2014}]{Gruberetal2014}
{Gruber} D.,  et~al., 2014, \mn@doi [\apjs] {10.1088/0067-0049/211/1/12}, \href
  {https://ui.adsabs.harvard.edu/abs/2014ApJS..211...12G} {211, 12}

\bibitem[\protect\citeauthoryear{{Hotan} et~al.,}{{Hotan}
  et~al.}{2014}]{Hotan2014}
{Hotan} A.~W.,  et~al., 2014, \mn@doi [\pasa] {10.1017/pasa.2014.36}, \href
  {https://ui.adsabs.harvard.edu/abs/2014PASA...31...41H} {31, e041}

\bibitem[\protect\citeauthoryear{{Inoue}}{{Inoue}}{2004}]{Inoue2004}
{Inoue} S.,  2004, \mn@doi [\mnras] {10.1111/j.1365-2966.2004.07359.x}, \href
  {https://ui.adsabs.harvard.edu/#abs/2004MNRAS.348..999I} {348, 999}

\bibitem[\protect\citeauthoryear{{Ioka}}{{Ioka}}{2003}]{Ioka2003}
{Ioka} K.,  2003, \mn@doi [\apj] {10.1086/380598}, \href
  {https://ui.adsabs.harvard.edu/#abs/2003ApJ...598L..79I} {598, L79}

\bibitem[\protect\citeauthoryear{{James} et~al.,}{{James}
  et~al.}{2019}]{James2019}
{James} C.~W.,  et~al., 2019, \mn@doi [\pasa] {10.1017/pasa.2019.1}, \href
  {http://ukads.nottingham.ac.uk/abs/2019PASA...36....9J} {36, e009}

\bibitem[\protect\citeauthoryear{{Kaplan} et~al.,}{{Kaplan}
  et~al.}{2015}]{Kaplan2015}
{Kaplan} D.~L.,  et~al., 2015, \mn@doi [\apjl] {10.1088/2041-8205/814/2/L25},
  \href {https://ui.adsabs.harvard.edu/abs/2015ApJ...814L..25K} {814, L25}

\bibitem[\protect\citeauthoryear{{Lorimer}, {Bailes}, {McLaughlin}, {Narkevic}
  \& {Crawford}}{{Lorimer} et~al.}{2007}]{Lorimeretal2007}
{Lorimer} D.~R.,  {Bailes} M.,  {McLaughlin} M.~A.,  {Narkevic} D.~J.,
  {Crawford} F.,  2007, \mn@doi [Science] {10.1126/science.1147532}, \href
  {http://adsabs.harvard.edu/abs/2007Sci...318..777L} {318, 777}

\bibitem[\protect\citeauthoryear{{Lyutikov}}{{Lyutikov}}{2018}]{Lyutikov2018}
{Lyutikov} M.,  2018, arXiv e-prints, \href
  {https://ui.adsabs.harvard.edu/abs/2018arXiv180910478L} {}

\bibitem[\protect\citeauthoryear{{Macquart}}{{Macquart}}{2007}]{Macquart2007}
{Macquart} J.-P.,  2007, \mn@doi [\apjl] {10.1086/513424}, \href
  {https://ui.adsabs.harvard.edu/abs/2007ApJ...658L...1M} {658, L1}

\bibitem[\protect\citeauthoryear{{Macquart} et~al.,}{{Macquart}
  et~al.}{2010}]{Macquart2010}
{Macquart} J.-P.,  et~al., 2010, \mn@doi [\pasa] {10.1071/AS09082}, \href
  {http://adsabs.harvard.edu/abs/2010PASA...27..272M} {27, 272}

\bibitem[\protect\citeauthoryear{{McConnell} et~al.,}{{McConnell}
  et~al.}{2016}]{McConnell2016}
{McConnell} D.,  et~al., 2016, \mn@doi [\pasa] {10.1017/pasa.2016.37}, \href
  {http://cdsads.u-strasbg.fr/abs/2016PASA...33...42M} {33, e042}

\bibitem[\protect\citeauthoryear{{Meegan} et~al.,}{{Meegan}
  et~al.}{2009}]{Meegan2009}
{Meegan} C.,  et~al., 2009, \mn@doi [\apj] {10.1088/0004-637X/702/1/791}, \href
  {https://ui.adsabs.harvard.edu/abs/2009ApJ...702..791M} {702, 791}

\bibitem[\protect\citeauthoryear{{Narayana Bhat} et~al.,}{{Narayana Bhat}
  et~al.}{2016}]{Narayana2016}
{Narayana Bhat} P.,  et~al., 2016, \mn@doi [\apjs]
  {10.3847/0067-0049/223/2/28}, \href
  {https://ui.adsabs.harvard.edu/abs/2016ApJS..223...28N} {223, 28}

\bibitem[\protect\citeauthoryear{{Palaniswamy}, {Wayth}, {Trott}, {McCallum},
  {Tingay}  \& {Reynolds}}{{Palaniswamy} et~al.}{2014}]{Palaniswamy2014}
{Palaniswamy} D.,  {Wayth} R.~B.,  {Trott} C.~M.,  {McCallum} J.~N.,  {Tingay}
  S.~J.,   {Reynolds} C.,  2014, \mn@doi [\apj] {10.1088/0004-637X/790/1/63},
  \href {https://ui.adsabs.harvard.edu/abs/2014ApJ...790...63P} {790, 63}

\bibitem[\protect\citeauthoryear{{Pshirkov} \& {Postnov}}{{Pshirkov} \&
  {Postnov}}{2010}]{PshirkovPostnov2010}
{Pshirkov} M.~S.,  {Postnov} K.~A.,  2010, \mn@doi [\apss]
  {10.1007/s10509-010-0395-x}, \href
  {https://ui.adsabs.harvard.edu/abs/2010Ap%26SS.330...13P} {330, 13}

\bibitem[\protect\citeauthoryear{{Ravi}}{{Ravi}}{2019}]{Ravi2019}
{Ravi} V.,  2019, \mn@doi [Nature Astronomy] {10.1038/s41550-019-0831-y}, \href
  {https://ui.adsabs.harvard.edu/abs/2019NatAs.tmp..405R} {p.~405}

\bibitem[\protect\citeauthoryear{{Ravi} \& {Lasky}}{{Ravi} \&
  {Lasky}}{2014}]{RaviLasky2014}
{Ravi} V.,  {Lasky} P.~D.,  2014, \mn@doi [\mnras] {10.1093/mnras/stu720},
  \href {https://ui.adsabs.harvard.edu/abs/2014MNRAS.441.2433R} {441, 2433}

\bibitem[\protect\citeauthoryear{{Rowlinson}, {O'Brien}, {Metzger}, {Tanvir}
  \& {Levan}}{{Rowlinson} et~al.}{2013}]{Rowlinson2013}
{Rowlinson} A.,  {O'Brien} P.~T.,  {Metzger} B.~D.,  {Tanvir} N.~R.,   {Levan}
  A.~J.,  2013, \mn@doi [\mnras] {10.1093/mnras/sts683}, \href
  {https://ui.adsabs.harvard.edu/abs/2013MNRAS.430.1061R} {430, 1061}

\bibitem[\protect\citeauthoryear{{Shannon} et~al.,}{{Shannon}
  et~al.}{2018}]{Shannon18}
{Shannon} R.~M.,  et~al., 2018, \nat, \href
  {http://adsabs.harvard.edu/abs/2018Natur.562..386S} {562, 386}

\bibitem[\protect\citeauthoryear{{Sokolowski} et~al.,}{{Sokolowski}
  et~al.}{2018}]{Sokolowski2018}
{Sokolowski} M.,  et~al., 2018, \mn@doi [\apj] {10.3847/2041-8213/aae58d},
  \href {https://ui.adsabs.harvard.edu/\#abs/2018ApJ...867L..12S} {867, L12}

\bibitem[\protect\citeauthoryear{{Staley} \& {Fender}}{{Staley} \&
  {Fender}}{2016}]{Staley2016}
{Staley} T.~D.,  {Fender} R.,  2016, preprint, \href
  {http://adsabs.harvard.edu/abs/2016arXiv160603735S} {} (\mn@eprint {arXiv}
  {1606.03735})

\bibitem[\protect\citeauthoryear{{Totani}}{{Totani}}{2013}]{Totani2013}
{Totani} T.,  2013, \mn@doi [\pasj] {10.1093/pasj/65.5.L12}, \href
  {https://ui.adsabs.harvard.edu/abs/2013PASJ...65L..12T} {65, L12}

\bibitem[\protect\citeauthoryear{{Usov} \& {Katz}}{{Usov} \&
  {Katz}}{2000}]{UsovKatz2000}
{Usov} V.~V.,  {Katz} J.~I.,  2000, \aap, \href
  {https://ui.adsabs.harvard.edu/abs/2000A%26A...364..655U} {364, 655}

\bibitem[\protect\citeauthoryear{{Wang}, {Yang}, {Wu}, {Dai}  \& {Wang}}{{Wang}
  et~al.}{2016}]{Wangetal2016}
{Wang} J.-S.,  {Yang} Y.-P.,  {Wu} X.-F.,  {Dai} Z.-G.,   {Wang} F.-Y.,  2016,
  \mn@doi [\apjl] {10.3847/2041-8205/822/1/L7}, \href
  {https://ui.adsabs.harvard.edu/abs/2016ApJ...822L...7W} {822, L7}

\bibitem[\protect\citeauthoryear{{Wang}, {Peng}, {Wu}  \& {Dai}}{{Wang}
  et~al.}{2018}]{Wangetal2018}
{Wang} J.-S.,  {Peng} F.-K.,  {Wu} K.,   {Dai} Z.-G.,  2018, \mn@doi [\apj]
  {10.3847/1538-4357/aae531}, \href
  {https://ui.adsabs.harvard.edu/abs/2018ApJ...868...19W} {868, 19}

\bibitem[\protect\citeauthoryear{{Zhang}}{{Zhang}}{2014}]{Zhang2014}
{Zhang} B.,  2014, \mn@doi [\apjl] {10.1088/2041-8205/780/2/L21}, \href
  {http://adsabs.harvard.edu/abs/2014ApJ...780L..21Z} {780, L21}

\makeatother
\end{thebibliography}

\bsp	
\label{lastpage}
\end{document}